 \documentstyle[prl,aps,multicol]{revtex}

\draft

\begin{document}

\title{ Theoretical estimates for the largest Lyapunov exponent
        of many-particle systems }

\author{ Ra\'ul O. Vallejos\thanks{{\rm e-mail: vallejos@cbpf.br}}
  and    Celia Anteneodo\thanks{{\rm e-mail: celia@cbpf.br   }} }

\address{Centro Brasileiro de Pesquisas F\'{\i}sicas,
         R. Dr. Xavier Sigaud 150, \\
         22290-180, Rio de Janeiro, Brazil}

\date{\today}

\maketitle

\begin{abstract}

The largest Lyapunov exponent of an ergodic Hamiltonian system is
the rate of exponential growth of the norm of a typical vector in
the tangent space. For an $N$-particle Hamiltonian system, with a
smooth Hamiltonian of the type $p^2 + {\cal V}(q)$, the evolution 
of tangent vectors is governed by the Hessian matrix ${\bf V}$ of 
the potential.
Ergodicity implies that the Lyapunov exponent is independent of 
initial conditions on the energy shell, which can then be chosen 
randomly according to the microcanonical distribution.
In this way a stochastic process ${\bf V}(t)$ is defined, and the
evolution equation for tangent vectors can now be seen as a stochastic
differential equation.
An equation for the evolution of the {\em average squared norm} of a 
tangent vector can be obtained using the standard theory 
in which the average propagator is written as a cummulant
expansion. We show that if cummulants higher than the second one are
discarded, the Lyapunov exponent can be obtained by diagonalizing
a small-dimension matrix, which, in some cases,
can be as small as $3 \! \times  \!3$.
In all cases the matrix elements of the propagator are expressed in
terms of correlation functions of the stochastic process.
We discuss the connection between our approach and an alternative
theory, the so-called geometric method.

\end{abstract}


\begin{multicols}{2}

\narrowtext

%%%%%%%%%%%%%%%%%%%%%%%%%%%%%%%%%%%%%%%%%%%%%%%%%%%%%%%%%%%%%%%%%%%%%%%%

\section{Introduction}
\label{sec:introduction}

%%%%%%%%%%%%%%%%%%%%%%%%%%%%%%%%%%%%%%%%%%%%%%%%%%%%%%%%%%%%%%%%%%%%%%%%

The largest Lyapunov exponent measures the sensitivity to initial 
conditions in a dynamical system. In low-dimensional models,
the Lyapunov exponent sets a limit to the prediction of the time 
evolution of a given state of the system.
In the high dimensional systems encountered in thermodynamics, 
one abandons from the start a description in terms of single 
(microscopic) states, and resorts to a statistical approach. 
In these cases a positive Lyapunov exponent
is usually welcome, as it is a necessary condition for the validity of the
Boltzmann-Gibbs scenario. A Lyapunov exponent that becomes null in the 
thermodynamic limit is a signal of anomalous behaviors.
For instance, metastable phases of some long-range interacting systems,
where the Lyapunov exponent vanishes in the large $N$ limit, exhibit
breakdown of ergodicity, anomalous diffusion, and non-Maxwell velocity
distributions \cite{metastable}.
An extension of standard statistical mechanics is 
required for the theoretical explanation of such phenomena \cite{tsallis}. 

In many-particle systems the Lyapunov exponent is also an indicator 
(order parameter) of phase transitions \cite{butera87,mehra97,barre01} 
and may be related to transport coefficients \cite{dorfman95,barnett96}.

In practice, for a given system, the largest Lyapunov exponent
must be obtained through numerical simulations, typically using the method 
developed by Benettin and collaborators \cite{benettin76} (a proof that 
this method gives the Lyapunov exponent of Oseledec's theorem 
\cite{oseledec68} can be found in Ref.~\cite{ershov98}). 
For some special cases, e.g., hard-sphere systems, the theory of Lyapunov 
exponents is remarkably developed \cite{dorfman99}. 
This contrasts with the situation in smooth Hamiltonian systems, where
comprehensive analytical estimates are scarce. In 1984, when studying 
two-dimensional billiards, Benettin made the first step towards the 
construction of random matrix methods for modeling {\em universal} 
features of Lyapunov spectra \cite{benettin84}. These methods were
further developed by other authors and applied to
many-particle smooth Hamiltonians \cite{rmt}.

To our knowledge, the first theory for estimating the largest Lyapunov 
exponent of a specific Hamiltonian system, 
and its dependence on the system parameters, was formulated by 
Pettini and coworkers some years ago \cite{geometric,casetti00}.
In this approach, the dynamics is geometrized by absorbing the force terms 
into a suitable metric, thus mapping the Hamiltonian problem onto a geodesic 
motion on a curved manifold.
After making the ``quasi-isotropy" approximation, 
the Hamilton equations for the tangent vectors become decoupled. 
As a consequence, the initial system of $2N$ differential equations has been 
reduced to only two equations. While the original problem was governed by 
the Hessian matrix of the potential, of size $N \! \times \! N$, 
the new (reduced) one is controlled by the Laplacian of the potential,
$\triangle {\cal V}(t)$, a scalar function of time. Thereafter, $\triangle 
{\cal V}(t)$ is treated as Gaussian white noise and the $2 \! \times \! 2$ 
system of differential equations is solved using the methods developed by 
van Kampen and others \cite{vankampen}. 
See Ref.~\cite{casetti00} for a review.

When applied to a Fermi-Pasta-Ulam chain \cite{pettini93}, the so-called
``geometric method" was extremely successful in reproducing the largest 
Lyapunov exponent over the entire energy range \cite{casetti00}. 
However, in other cases the agreement is not so good. For instance,
in a chain of rotators with first-neighbor (bounded) interactions, the 
method works well only in the low and high energy regimes, where the 
dynamics is weakly chaotic (integrable in the limits $E \to 0,\infty$). 
In the intermediate region of stronger chaos, the theory has to be 
amended to obtain a good agreement with simulations \cite{casetti00}. 
This and other examples \cite{barre01,latora99} raise several questions 
concerning the domain of validity of the theory.
What is the nature of the quasi-isotropic approximation? Or, what are the
parameters that control the quality of the estimates of the theory? 
Is the geometric method perturbative? If so, what 
are the next leading corrections?

In this paper we present an alternative theoretical approach in which 
the validity domains of the successive approximations can be precisely 
delimited. The basic idea is to employ van Kampen's methods \cite{vankampen} 
to solve the {\em original} system of $2N$ differential equations for 
the evolution of tangent vectors. By applying this scheme to a three 
dimensional dilute gas, Barnett et al. \cite{barnett96} established a link 
between the Lyapunov exponent and the self-diffusion coefficient 
(see also \cite{torcini99}). We show that this approach can be extended
to reach other systems, like Fermi-Pasta-Ulam chains and lattices of
classical spins, either with short or long-range interactions. 
By doing so, we shall settle down a connection with the results
of the geometric method and suggest some answers to the above-mentioned
questions.

The paper has been organized as follows.
Section~\ref{sec:theory} presents the theory that leads to an estimate of
the largest Lyapunov exponent. This is a perturbative theory, which rests 
on a cummulant expansion.
We argue that the general (perturbative) solution can be obtained by 
diagonalizing a small-dimension matrix.
In Sect.~\ref{sec:isotropic} we analyze an approximation that reduces the 
problem to diagonalizing a $3 \! \times \! 3$ matrix. 
Section~\ref{sec:examples} discusses some examples which illustrate the 
working of the theory.
The connection between our results and those obtained by the geometric
method is discussed in Sect.~\ref{sec:geometric}. 
Finally, Sect.~\ref{sec:conclusions} contains the concluding remarks.

%%%%%%%%%%%%%%%%%%%%%%%%%%%%%%%%%%%%%%%%%%%%%%%%%%%%%%%%%%%%%%%%%%%%%%%%

 \section{Theory}
 \label{sec:theory}

%%%%%%%%%%%%%%%%%%%%%%%%%%%%%%%%%%%%%%%%%%%%%%%%%%%%%%%%%%%%%%%%%%%%%%%%

The theory we present in this section can, in principle,
be applied to any smooth Hamiltonian system. For simplicity, and for the
sake of comparisons with the geometric method,
we restrict ourselves to ``natural" Hamiltonians,
\begin{equation}
\label{ham}
{\cal H} = \sum_{i=1}^N \frac{ p^2_i}{2m} + {\cal V}(q_1,\ldots,q_N),
\end{equation}
where $q_i$ and $p_i$, are conjugate position-momentum coordinates.
Other Hamiltonians can be considered but they may require modifications
of the theory.

The Hamilton equations can be written in the compact form
\begin{equation} \label{motionx}
\dot{x} \;=\;  {\bf J}\frac{\partial {\cal H}}{\partial x}
 \; ,
\end{equation}
where we have introduced the $2N$-dimensional column vector $x$,
$x=(q_1,\ldots,q_N,p_1,\ldots, p_N)^T$,
the superscript meaning ``transposed", and the
symplectic matrix $\bf J$,
\begin{equation}
 {\bf J} \;=\; \left( \matrix{    0      & \openone \cr
                               -\openone &    0 }\right)
 \; ,
\end{equation}
with $\openone$ the $N\times N$ identity matrix.
Differentiating the Hamilton equations, one obtains the evolution 
equations for tangent vectors $\xi=(\delta q_1,\ldots,\delta q_N,
\delta p_1,\ldots, \delta p_N)^T$,
\begin{equation}
\label{tangent0}
\dot \xi = {\bf A}(t)\,\xi.
\end{equation}
For a Hamiltonian of the special form (\ref{ham}), and setting
$m=1$, the operator $\bf A$ has the simple structure
\begin{equation} \label{A}
{\bf A}(t) =
\left( \matrix{    0          & \openone \cr
                -{\bf V}(t)   &    0          }\right) \; .
\end{equation}
Here $\bf V$ is the Hessian matrix of the potential ${\cal V}$, namely
\begin{equation} \label{V}
V_{ij} \;=\;  \frac{\partial^2{\cal V}}{\partial q_i \partial q_j} \;.
\end{equation}
Once initial conditions $x_0$ and $\xi_0$ have been specified,
Eqs.~(\ref{motionx})
and (\ref{tangent0})
allow one to find the Lyapunov exponent $\lambda$ by calculating the limit
\cite{benettin76}
\begin{equation}
\label{defliapunov}
\lambda = \lim_{t \to \infty} \frac{1}{2t} \ln | \xi (t; x_0,\xi_0)|^2   \;
.
\end{equation}

We will assume that for any initial condition $x_0$, the phase-space
trajectory $x(t;x_0)$ is ergodic on its energy shell.
This implies that $\lambda$ is independent of initial conditions
$x_0$, which can then be chosen randomly according to the microcanonical
distribution. There will also be no dependence on initial tangent vectors,
because if $\xi_0$ is also chosen randomly, it will have a non-zero 
component along the most expanding direction.

If the corrections to the exponential law in Eq.~(\ref{defliapunov})
go to zero fast enough as $t \to \infty$, one can also write
\begin{equation}
\label{defliapunov2}
\left \langle | \xi (t; x_0,\xi_0)|^2 \right \rangle  
\propto e^{2 \lambda t} \; ,
\end{equation}
brackets meaning microcanonical averages over $x_0$.
We will prefer the estimate of Eq.~(\ref{defliapunov2})
because the averaging procedure is crucial for finding an analytical
expression for the Lyapunov exponent.
In case of doubt, the equality of the exponents defined by
Eqs.~(\ref{defliapunov}) and (\ref{defliapunov2}) can be tested 
numerically, e.g., using the data generated by Benettin's algorithm.

By letting $x_0$ be a random variable,
a stochastic process ${\bf V}(t;x_0)$ is defined, and Eq.~(\ref{tangent0})
can be thought as a stochastic differential equation.
However, the quantity we are interested in is the square of 
the norm of $\xi$, which can be written as the trace of the
``density matrix" $\xi \xi^T$. Thus, we must focus on the equation for the
evolution of $\xi \xi^T$:
\begin{equation} \label{vK0}
\frac{\rm d}{{\rm d} t} (\xi\xi^T) = {\bf A}\xi\xi^T \;+\; \xi\xi^T{\bf A}^T
\;\equiv\;
\widehat{\bf A}\,\xi\xi^T,
\end{equation}
the rightmost identity defining the linear superoperator $\widehat{\bf A}$.
Except for the fact that we must deal now with a superoperator,
Eq.~(\ref{vK0}) is not different from Eq.~(\ref{tangent0}) 
and can be handled with the same techniques.
For the purpose of the perturbative approximations that will follow, the
operator $\widehat{\bf A}$ is split into two parts
\begin{equation}
\widehat{\bf A} \,=\,\widehat{\bf A}_0 \,+\,\widehat{\bf A}_1(t) \; ,
\end{equation}
where $\widehat{\bf A}_0$ corresponds to the evolution in the absence of
interactions. In our case $\widehat{\bf A}_0$ and $\widehat{\bf A}_1$
are associated with
\begin{equation} \label{a0a1}
{\bf A}_0 =  \left( \matrix{    0    &  \openone \cr
                                0    &  0             }\right)
\;\;\;\;\;\mbox{and}\;\;\;\;\;
{\bf A}_1 =  \left( \matrix{    0           &  0 \cr
                                -{\bf V}(t) &  0      }\right) \, ,
\end{equation}
respectively.
Whenever ${\bf A}_1(t)$ is small (in a sense that will be discussed below),
it is possible to manipulate Eq.~(\ref{vK0}) to derive an explicit 
expression for the evolution of the {\em average} of $\xi\xi^T$.
A clear exposition of this derivation, together with a very detailed
discussion of its domain of validity has been given by van Kampen 
\cite{vankampen}.
We just outline the basic steps:
(a) Rewrite Eq.~(\ref{vK0}) in the interaction representation associated
    with $\widehat{\bf A}_0$.
(b) Write the propagator as a time ordered exponential.
(c) Expand its average in a series of cummulants. 
(d) Go back to the original representation.
    The final result is:
\begin{equation}
\label{solution}
  \langle {\xi\xi^T}\rangle(t) =
e^{ t\widehat{\bf \Lambda}}\;{\xi_0\xi_0^T}\;,
\end{equation}
where $\widehat{\bf \Lambda}$ is a time-independent superoperator given by
the perturbative expansion:
\begin{eqnarray} \label{expansion} \nonumber
&&\widehat{\bf \Lambda} \; \equiv \;
 \widehat{\bf A}_0 \;+\;
   \langle \widehat{\bf A}_1 \rangle \\ \label{final}
&&+
\int_0^\infty {\rm d} \tau
\left\langle\delta \widehat{\bf A}_1(t)\, e^{ \tau \widehat{\bf
A}_0 } \,
  \delta\widehat{\bf A}_1(t-\tau) \,e^{ -\tau \widehat{\bf
A}_0 }
\right\rangle + \cdots \; ,
\end{eqnarray}
with
\begin{equation}
\delta \widehat{\bf A}_1(t)= \widehat{\bf A}_1(t) -\langle \widehat{\bf A}_1
\rangle \; .
\end{equation}
Let $L_{\rm \scriptsize max}$ be the  eigenvalue of $\widehat{\bf \Lambda}$
which has the largest real part. 
We find that the largest Lyapunov exponent $\lambda$ is related to the real part of 
$L_{\rm \scriptsize max}$:
\begin{equation}
\lambda = \mbox{$\frac{1}{2}$} \, \mbox{Re} \, \left( L_{\rm \scriptsize max} \right) \;.
\end{equation}

In Eq.~(\ref{expansion}) we give explicitly only the first two cummulants,
the dots stand for third cummulants and higher order ones. 
The perturbative parameter can be understood as the product of two quantities.
The first one, let's call it $\sigma$, characterizes the amplitude of the 
fluctuations of $\delta \widehat{\bf A}_1(t)$.
The second, $\tau_c$, is a typical (the largest relevant) correlation 
time of $\delta \widehat{\bf A}_1(t)$.
Thus, the second cummulant is of the order of $\sigma^2 \tau_c$, the third
one is of the order of $\sigma^3 \tau^2_c$, and so on. 
If all cummulants were summed up, Eq.~(\ref{solution}) would be exact 
in the long-time regime $t \gg \tau_c$ \cite{vankampen}.

From now on, we restrict our analysis to the propagator 
$\widehat{\bf \Lambda}$ truncated
at the second order, i.e., Eq.~(\ref{expansion}) without the dots.
This approximation will be better the smaller $\sigma \tau_c$. 
However, if $\widehat{\bf A}_1(t)$ is not far from a Gaussian process,
the validity of the second order approximation may extend outside the
perturbative region $\sigma \tau_c \ll 1$.
In the exceptional case that $\widehat{\bf A}_1(t)$ is a Gaussian process,
cummulants higher than the second one will be strictly zero  
and the truncation will introduce no error.

To proceed further one needs the matrix of $\widehat{\bf \Lambda}$ 
in some basis.
So, let us calculate $\widehat{\bf \Lambda} {\bf M}$, with $\bf M$ a
symmetric matrix (it is easy to see that the truncation has not 
spoiled the symmetry of the density matrix).
First notice that the exponentials of $\widehat{\bf A}_0$ represent no
problem as they are finite polinomials,
\begin{equation}
e^{ \tau \widehat{\bf A}_0} {\bf Q}\;=
\;[\openone+\tau {\bf A}_0]\,{\bf Q}\,[\openone+\tau {\bf A}_0^T],
\end{equation}
for any matrix $\bf Q$. Inserting this expression into (\ref{final})
we arrive at
\begin{eqnarray} \nonumber
&&\widehat{\bf \Lambda} \,{ \bf M} \;=\;
 ({\bf A}_0 + \langle {\bf A}_1\rangle)\,{\bf M} \\ \nonumber
&&+\;
\int_0^\infty {\rm d} \tau
\left\langle\delta {\bf A}_1(t)\,
[\delta\widetilde{\bf A}_1(t-\tau) {\bf M} \,+\,
{\bf M}\delta\widetilde{\bf A}_1^T(t-\tau)] \right\rangle \\
&&+\;(\cdots)^T \; , \label{lambdaM}
\end{eqnarray}
where $(\cdots)^T$ means ``the previous terms transposed", and
\begin{equation}
\delta\widetilde{\bf A}_1(t-\tau)\;=\;[\openone+\tau {\bf A}_0]\;
\delta {\bf A}_1(t-\tau)\;[\openone-\tau {\bf A}_0]\, .
\end{equation}
Substituting (\ref{a0a1}) into  (\ref{lambdaM}) we arrive at
the final result of the second-order perturbative approach
\begin{eqnarray} \label{lambdaM2} \nonumber
&&\widehat{\bf \Lambda}  {\bf M}  =
\left( \matrix{    0                           &  \openone \cr
                 -\langle {\bf V} \rangle     &  0             }
\right) {\bf M}  \\ [5mm] \nonumber && +
\int_{0}^\infty d\tau \left( \matrix{    0     &  0  \cr
                                      \tau     &  -\tau^2  }
\right)
\left( \matrix{  \left\langle \delta {\bf V}\delta {\bf V'} \right\rangle  &
0 \cr
                            0   &  \left\langle \delta {\bf V}\delta {\bf
V'} \right\rangle  }
\right) {\bf M}   \\[5mm] \nonumber && +
\int_{0}^\infty d\tau \left\langle
\left( \matrix{    0                     &  0 \cr
                 \delta {\bf V}          &  0     }
\right)  {\bf M}
\left( \matrix{  \delta {\bf V'}   &       0    \cr
                   0               &   \delta {\bf V'}   }
\right)
\right\rangle
\left( \matrix{   \tau     &   \openone  \cr
                 -\tau^2   &  -\tau       }
\right)
\\[5mm] &&+ (\cdots)^T \; .
\end{eqnarray}
To abbreviate the notation, we have written $\tau^{n}$ instead of
$\tau^{n} \openone$;  $\delta {\bf V}$ and  $\delta {\bf V'}$ 
substitute $\delta {\bf V}(t)$ and  $\delta {\bf V}(t-\tau)$, respectively.

The largest Lyapunov exponent is buried into Eq.~(\ref{lambdaM2}).
To get an explicit expression one must diagonalize the matrix of
$\widehat{\bf \Lambda}$. The outcome will be $\lambda$ as a function of the
first two cummulants of the stochastic process ${\bf V}(t)$, i.e., averages
and (integrated) two-times correlation functions:
\begin{equation}
\langle V_{ij} \rangle \; ;  \;\;\;
\int_0^\infty d\tau \, \tau^n \,
\langle \delta V_{ij}(0)\delta V_{kl}(\tau) \rangle \; , \;\;\; n=0,1,2 .
\end{equation}
At first sight it may be thought that, as $\widehat{\bf \Lambda}$ is a
superoperator, the matrix one should diagonalize is of the order of
$N^2 \! \times \! N^2$, then straightforward diagonalization would be
out of the question for large $N$.
Notwithstanding, $\widehat{\bf \Lambda}$ is an averaged object, and, as
such, it possesses some symmetries which can be exploited to reduce the
dimensionality of the problem to tractable levels, say $N \! \times \! N$.
This will be illustrated with an example in Sect.~\ref{sec:examples}.
So, if desired, the largest Lyapunov
exponent could be found by numerical diagonalization, at least for systems
with $N \approx 1000$ degrees of freedom (provided one can estimate the
correlation functions).

An alternative to exact diagonalization is approximate diagonalization,
i.e., the diagonalization of the restriction of $\widehat{\bf \Lambda}$ 
to some small-dimension subspace. Notice that the problem we are dealing 
with is not very different from finding the ground state energy of a 
quantum system. However, in the quantum problem, the operator one must 
diagonalize is Hermitian, and it is well known that diagonalization in 
a truncated basis produces an upper bound to the exact ground-state energy. 
It frequently happens that this bound is close to the exact result, 
even if the ground-state wavefunction is not. In spite of the operator 
$\widehat{\bf \Lambda}$ not being
Hermitian (see Sect.~\ref{sec:isotropic}) we still expect that
diagonalization in a small basis will give a lower bound for the 
Lyapunov exponent. If the basis is  suitably chosen, this estimate 
may be close to the result of the exact diagonalization.

To proceed with the construction of a basis for $\widehat{\bf \Lambda}$, 
we take advantage of the fact that $\lambda$ is independent of $\xi_0$,
and simplify Eq.~(\ref{solution}) further by averaging over an 
orthonormal set of initial tangent vectors, obtaining
\begin{equation}
\langle \langle {\xi\xi^T} \rangle \rangle (t) =
\frac{1}{2N}\;e^{t\widehat{\bf \Lambda}}\;\openone \;.
\end{equation}
This second averaging allows us to consider, 
instead of $\widehat{\bf \Lambda}$
itself, the restriction of $\widehat{\bf \Lambda}$ to the subspace spanned
by the matrices $\widehat{\bf \Lambda}^k\openone$, $k=0,1,2,\ldots$.
A look at the first terms of these sequence gives a hint for constructing
an appropriate basis. The first term is the identity, the second one is
\begin{eqnarray} \label{lambda1}
\widehat{\bf \Lambda} \openone & = &
\left( \matrix{    0                                   &  \openone -
\langle {\bf V} \rangle \cr
                  \openone - \langle {\bf V} \rangle   &  0
              }\right) \nonumber \\   & & +
2 \int_{0}^\infty d\tau
\left( \matrix{    0                                   &  \tau
\langle \delta {\bf V} \delta{\bf V}'\rangle \cr
                  \tau \langle \delta{\bf V} \delta{\bf V}'\rangle &  (1-\tau^2)
\langle \delta{\bf V} \delta{\bf V}'\rangle
              }\right)
\; ,
\end{eqnarray}
and so on.

%%%%%%%%%%%%%%%%%%%%%%%%%%%%%%%%%%%%%%%%%%%%%%%%%%%%%%%%%%%%%%

\section{The isotropic approximation}
\label{sec:isotropic}

%%%%%%%%%%%%%%%%%%%%%%%%%%%%%%%%%%%%%%%%%%%%%%%%%%%%%%%%%%%%%%

Typically, the diagonal elements of ${\bf V}(t)$ will be larger 
than the off-diagonal ones. This is evident in the case of 
translational invariance, where one has the property
\begin{equation}
\label{viisumvij}
V_{ii}=-\sum_{j\neq i} V_{ij} \;.
\end{equation}
Introducing a matrix $\bf Y$ having all entries equal to one, i.e.,
\begin{equation} \label{ypsilum}
{\bf Y}_{ij} = 1, \;\;\;\; \forall i,j  \; .
\end{equation}
we can rewrite Eq.~(\ref{viisumvij}) as
\begin{equation} \label{yvvy}
{\bf Y}{\bf V} = {\bf V}{\bf Y} = 0 \; .
\end{equation}
Then it is clear that Eq.~(\ref{viisumvij}) is also satisfied by
$\langle {\bf V} \rangle$,
$\langle \delta {\bf V} \delta {\bf V}' \rangle$, 
and by the higher moments of ${\bf V}$ that will appear in the blocks 
of $\widehat{\bf \Lambda}^k\openone$, for $k>1$. 
So, in a first (crude) approximation
one may be tempted to discard the off-diagonal elements of the moments of 
${\bf V}$. If we also assume that all coordinates
$q_i$ are statistically equivalent, and remind that
the matrices $\widehat{\bf \Lambda}^k\openone$ are symmetric, we arrive at
the simplest approximation for diagonalizing $\widehat{\bf \Lambda}$. 
We call this approximation ``isotropic'',
and it consists in restricting $\widehat{\bf \Lambda}$ to the subspace
spanned by the following three matrices:
\begin{equation} \label{i1i2i3}
{\bf I}_1=\left( \matrix{  \openone          &       0    \cr
                       0               &       0  } \right), \;
{\bf I}_2=\left( \matrix{    0               &       0    \cr
                       0               &   \openone  } \right), \;
{\bf I}_3=\left( \matrix{    0               &   \openone    \cr
                     \openone          &       0  } \right).
\end{equation}
These matrices are mutually orthogonal with respect to the
standard Euclidean scalar product, i.e.,
\begin{equation}
{\rm Tr} ({\bf I}_i{\bf I}_j^T) \propto \delta_{ij} \;.
\end{equation}
Then the matrix elements of $\widehat{\bf \Lambda}$ with respect to the
basis
$\{ {\bf I}_1,{\bf I}_2, {\bf I}_3\}$ are
\begin{equation}
\Lambda_{ij}^{II} \;=\; \frac
{ {\rm Tr}([\widehat{\bf \Lambda} {\bf I}_j] {\bf I}_i^T )}
{ {\rm Tr} ({\bf I}_i {\bf I}_i^T) }   \; .
\end{equation}
Using Eq.~(\ref{lambdaM2}) and skipping some simple algebra, 
we arrive at the 3$\times$3 matrix
\begin{equation}
\label{iso}
{\bf \Lambda}^{II}\;=\;
\left( \matrix{ 0                       &  0                      &  2
\cr
                2\sigma^2 \tau_c^{(1)}  & -2\sigma^2 \tau_c^{(3)} & -2\mu
\cr
          -\mu+ 2\sigma^2 \tau_c^{(2)}  &  1
                     & -2\sigma^2 \tau_c^{(3)} \cr
  } \right)                                \; ,
\end{equation}
with the definitions
\begin{eqnarray}
\mu            &=& \frac{1}{N} {\rm Tr } \langle {\bf V} \rangle          \; ,   \\
\sigma^2       &=& \frac{1}{N} {\rm Tr } \langle \left( \delta {\bf V} \right)^2 \rangle        \; ,   \\
\tau_c^{(k+1)} &=& \int_0^{\infty} d\tau \, \tau^k f(\tau)  \label{tau} \; ,
\end{eqnarray}
where we have introduced the normalized correlation function $f(\tau)$
\begin{eqnarray}
f(\tau) & = & \frac{1}{N \sigma^2} {\rm Tr }      \langle \delta {\bf V}(0)
\delta {\bf V}(\tau) \rangle  \nonumber \\
        & = & \frac{1}{N \sigma^2} \sum_{i,j=1}^N \langle \delta  V_{ij}(0)
\delta  V_{ij}(\tau) \rangle     \;.
\end{eqnarray}
It is evident from Eq.~(\ref{iso}) that the operator $\hat{\bf \Lambda}$ is
not Hermitian. Normalization
of the basis ${\bf I}_j$ will not make ${\bf \Lambda}^{II}$ symmetric.

In the isotropic approximation the Lyapunov exponent is expressed in terms
of the set of four parameters
$\mu$ and $\sigma^2 \tau_c^{(k+1)}$, $k=0,1,2$.
The parameters $\mu$ and $\sigma$ are, respectively, the mean and variance
of the stochastic
process ${\bf V}(t)$, and can in principle be obtained analytically by
calculating the corresponding
microcanonical averages.
(In practice the calculations can be done in the canonical ensemble, and
then connected with the microcanonical results by the formula of
Lebowitz, Percus and Verlet \cite{lebowitz67}.)

The characteristic time $\tau_c^{(1)}$ is naturally interpreted 
as the correlation time of the process ${\bf V}(t)$. 
Its calculation requires the knowledge of the autocorrelation functions
of $V_{ij}(t)$, which are system dependent. Moreover, the 
correlation functions of a given system will in general depend on energy. 
However, if the functional form of the correlation function 
$f(\tau)$ is known (or conjectured), its parameters can also be 
calculated as thermal averages.
For instance, if $f(\tau)$ is approximately Gaussian, 
\begin{equation}
f(\tau) \approx  e^{-\gamma \tau^2} \;,
\end{equation}
the expansion of
$ \langle  {\bf V}(0) {\bf V}(\tau) \rangle $
around $\tau=0$ gives an explicit formula for the correlation
time, namely
\begin{equation}
\frac{1}{\tau_c^{(1)}}=
\left[ \frac{2}{\pi \sigma^2 N}
\mbox{Tr} \left \langle \left( \frac{d{\bf V}}{dt} \right)^2 \right \rangle
\right]^{1/2} \; .
\end{equation}
In this case $\tau_c^{(2)}$ and $\tau_c^{(3)}$ are trivially related to
$\tau_c^{(1)}$:
\begin{eqnarray}
\tau_c^{(2)} &=& \frac{2}{\pi} \left[ \tau_c^{(1)} \right]^2 \; ,   \\
\tau_c^{(3)} &=& \frac{2}{\pi} \left[ \tau_c^{(1)} \right]^3 \; .
\end{eqnarray}

A purely numerical calculation of $\tau_c^{(k)}$ may be very difficult, 
if not impossible, because correlation functions estimated from 
finite-length time series usually fail to damp as expected \cite{jenkins68}.
Perhaps, a more sensible approach to the estimation of $\tau_c^{(k)}$ should
start with a numerical study of the correlation functions;
then a functional form for $f(\tau)$ could be proposed, based on the
short-time behavior of the numerical correlation functions; finally, 
the parameters defining $f(\tau)$ would be calculated as suitable thermal 
averages. An alternative, more powerful approach, involves the use of 
``memory functions" \cite{memory}: They are related in a one-to-one
way to correlation functions and seem to be more amenable to
simple approximations (see, e.g., Ref.~\cite{sharma99}).

%%%%%%%%%%%%%%%%%%%%%%%%%%%%%%%%%%%%%%%%

\section{Examples}
\label{sec:examples}

In this section we analize the application of the perturbative 
theory of Sect.~\ref{sec:theory} to some simple models. 
We remark that, in principle, the theory is expected to be
successful only in regimes where the Lyapunov exponent is very small.
All the systems considered below exhibit regimes with vanishingly small 
Lyapunov exponents. It is understood that our discussion will
be restricted to such regimes.

%%%%%%%%%%%%%%%%%%%%%%%%%%%%%%%%%%%%%%%%
%
\subsection{Mean field XY-Hamiltonian}
%
%%%%%%%%%%%%%%%%%%%%%%%%%%%%%%%%%%%%%%%%

Let us begin by analizing one special case in which the isotropic
approximation of Sect.~\ref{sec:isotropic} is exact. 
Consider the one-dimensional Hamiltonian  \cite{antoni95,latora98,yamaguchi96}
\begin{equation}
H_1  =  \frac{1}{2 } \sum_{i=1  }^N L_{i}^{2} +
        \frac{1}{2N} \sum_{i,j=1}^N [1-\cos(\theta_{i}-\theta_{j})] \; .
\end{equation}
This is the so-called mean-field $XY$-Hamiltonian.
It represents a lattice of classical spins with infinite-range interactions.
Each rotator is restricted to the unit circle and it is therefore described
by an angle $0 < \theta_i \le 2\pi$ and its conjugate angular momentum $L_i$, 
with $i=1,\ldots,N$. At the critical energy $E_c=3N/4$ there is a second-order 
phase transition separating a disordered regime ($E>E_c$) from an ordered one
($E<E_c$). 

In both limits $E \to 0,\infty$ the Lyapunov exponent goes to zero.
For a fixed energy $E>E_c$, $\lambda$ also goes to zero when $N\to \infty$. 
This behaviour has also been observed in a metastable disordered phase 
with $E<E_c$.
The perturbative approach should be a good approximation in these regimes.
Moreover, we argue that the infinite-range interactions justify the isotropic 
approximation. 

All single-particle averages are equal, and, given that the forces are 
independent of the distances between spins, all two-particle averages 
must also be equal. So, one has
\begin{eqnarray} \label{viivij}
\langle V_{ii} \rangle & = & c_1,   \;\;\;\;    \forall i \; , \\
\langle V_{ij} \rangle & = & c_2,   \;\;\;\;   \forall i \neq j \; .
\end{eqnarray}
Notice that translational symmetry, Eq.~(\ref{viisumvij}), implies that
\begin{equation}
 c_2 = -\frac{c_1}{N-1} \; . 
\end{equation}
This is the reason why the isotropic approximation will work in this 
case, i.e., off-diagonal matrix elements are indeed smaller than 
diagonal ones \cite{latora99}. But let us keep the discussion 
quantitave, and rewrite Eq.~(\ref{viivij}) as 
\begin{equation}
\langle {\bf V} \rangle = c_1 \openone + c_2 ({\bf Y}-\openone) \;,
\end{equation}
with ${\bf Y}$ defined in  Eq.~(\ref{ypsilum}).
Using the time-reversal symmetry of the stochastic process ${\bf V}(t)$,
one can also show that
\begin{equation}
\langle \delta {\bf V}(0) \delta {\bf V}(\tau) \rangle = c'_1 \openone + c'_2
({\bf Y}-\openone) \;.
\end{equation}
Then all blocks of $\widehat{\bf \Lambda} \openone$ [Eq.~(\ref{lambda1})] 
belong to the subspace spanned by $\openone$ and ${\bf Y}$. 
Taking into account Eq.~(\ref{yvvy}) and
\begin{equation}
{\bf Y}^2 =  N {\bf Y} \; ,
\end{equation}
we conclude that the blocks of all the sequence $\widehat{\bf \Lambda}^k \openone$
belong to the subspace $\{\openone,{\bf Y}\}$.
Thus, the relevant subspace for diagonalizing $\widehat{\bf\Lambda}$
is six-dimensional.
It is spanned by
${\bf I}_1,{\bf I}_2,{\bf I}_3$ [Eq.~(\ref{i1i2i3})] and
${\bf Y}_1,{\bf Y}_2,{\bf Y}_3$,
with the definitions
\begin{equation}
{\bf Y}_1=\left( \matrix{  {\bf Y}           &       0
\cr
                             0               &       0     } \right), \;
{\bf Y}_2=\left( \matrix{    0               &       0
\cr
                             0               &   {\bf Y}  } \right), \;
{\bf Y}_3=\left( \matrix{    0               &   {\bf Y}                 \cr
                           {\bf Y}           &       0     } \right) \; .
\end{equation}
However, it can be shown (see Appendix) that the largest
eigenvalue of the corresponding 6$\times$6 matrix coincides with that of the
isotropic 3$\times$3 matrix, up to corrections of order $1/N$. 
In this way we have proven the validity of the isotropic 3$\times$3 approximation 
for one-dimensional systems with infinite-range forces.

%%%%%%%%%%%%%%%%%%%%%%%%%%%%%%%
%
\subsection{Dilute gases}
%
%%%%%%%%%%%%%%%%%%%%%%%%%%%%%%%

Consider now a one-dimensional gas with Hamiltonian
\begin{equation}
H_2 = \frac{1}{2} \sum_{i=1}^N p_{i}^{2} + \sum_{i,j=1}^N \nu(q_i-q_j) \; ,
\end{equation}
where $q$ and $p$ are linear coordinates, and we assume that the 
potential $\nu$ is bounded.
For large enough energies, this system is disordered and weakly chaotic. 
All particles, and all {\em pairs} of particles, are statistically 
equivalent. Then this problem is formally equivalent to the infinite 
range $XY$-Hamiltonian: The isotropic 3$\times$3 approximation
becomes exact. 

Of course, the statistical equivalence also holds for a dilute 
three-dimensional gas with short-range interactions. 
In this case, Barnett {\em et al.} have shown that the
largest Lyapunov exponent is found by diagonalizing a 4$\times$4 matrix 
\cite{barnett96}.

%%%%%%%%%%%%%%%%%%%%%%%%%%%%%%%
%
\subsection{$\alpha XY$-Hamiltonian}
%
%%%%%%%%%%%%%%%%%%%%%%%%%%%%%%%

There are cases in which no strong reasons exist to believe that
the isotropic approximation will work satisfactorily. Consider, for
instance, the arbitrary-range analog of the $XY$-Hamiltonian 
\cite{anteneodo98,tamarit00}:
\begin{equation}
H_3 = \frac{1}{2} \sum_{i=1}^N L_{i}^{2} +
                      \frac{1}{2 \widetilde{N}} \sum_{i,j=1 \;(i \neq j)}^N
                      \frac{1-\cos(\theta_{i}-\theta_{j})}{r_{ij}^\alpha} \;
.
\end{equation}
The parameter $\alpha$ sets the range of the interactions: $\alpha=0$
recovers the mean-field case, $\alpha=\infty$ corresponds to 
first-neighbors couplings.
The prefactor $\widetilde{N}$ (a function of $N$ and $\alpha$) is included
to make the system ``pseudo-extensive" \cite{tsallis95}.
Periodic boundary conditions are assumed, and $r_{ij}$ is the minimum
between  $|i-j|$ and $N-|i-j|$. 
For any value of $\alpha$  there exist
(i) a low-energy regime of harmonic oscillators weakly coupled by
non-linear forces and
(ii) a high-energy disordered phase where the spins rotate almost freely. 
We expect our theory to produce good estimates for the largest Lyapunov
exponent in both low- and high-energy regimes.

If the forces are not infinite-range but just long-ranged, the isotropic
approximation will still give good estimates in weakly chaotic regimes. 
Evidence supporting
this statement can be found in Refs.~\cite{firpo01} (geometric method) and
\cite{anteneodo01} (random matrix approach), where some kind of ``isotropic"
approximations were used to derive scaling laws for $\lambda$ 
in the high-energy regime,
in good agreement with numerical simulations \cite{anteneodo98,campa01}.

For $\alpha$ not too small, it may be necessary to improve the isotropic
approximation by diagonalizing $\widehat{\bf \Lambda}$ in a larger basis.
In this case, the statistical equivalence holds for all pairs of particles 
separated by the same distance. This means that the blocks of
$\widehat{\bf \Lambda}^k\openone$ are symmetric and cyclical,
i.e., the matrix elements only depend on the distance $r_{ij}$.
A basis can be constructed starting from the $N$$\times$$N$ matrix
$\bf S$ of a cyclical shift:
\begin{equation}
S_{ij} = \delta_{i,j+1} \; ,
\end{equation}
where it is understood that $j+1$ must be taken modulo $N$. Then the set
of symmetrical matrices
\begin{equation}
{\bf \Sigma}_k \equiv {\bf S}^k + {\bf S}^{-k}\;, \;\;\;  0 \le k \le N/2 \;
,
\end{equation}
is a basis for the blocks of $\widehat{\bf \Lambda}^m\openone$. A suitable
basis for diagonalizing $\widehat {\bf \Lambda}$ is the set
\begin{equation}
\label{s1s2s3}
\left( \matrix{ {\bf \Sigma}_i     &       0              \cr
                 0                 &       0              }\right), \;
\left( \matrix{  0                 &       0              \cr
                 0                 &      {\bf \Sigma}_j  }\right), \;
\left( \matrix{  0                 &      {\bf \Sigma}_k  \cr
                 {\bf \Sigma}_k    &       0              }\right)  \; ,
\end{equation}
with $0 \le i,j,k \le N/2$. The length of this basis is 3$N$/2.
Notwithstanding, we expect that a small subset of this basis will
be enough to get a satisfactory convergence to the largest eigenvalue
of $\widehat{\bf \Lambda}$.
Even in the worst case of no truncation at all,
numerical diagonalization is possible for relatively large systems.

%%%%%%%%%%%%%%%%%%%%%%%%%%%%%%%%%%%%%%%%%%%%%

\section{Connection with the geometric method}
\label{sec:geometric}

%%%%%%%%%%%%%%%%%%%%%%%%%%%%%%%%%%%%%%%%%%%%%

In Sect.~\ref{sec:isotropic} we motivated the isotropic approximation by arguing
that, in a first approach, one can neglect the off-diagonal matrix elements
of the blocks of $\hat{\bf \Lambda}^k\openone$. Then, in Sect.~\ref{sec:examples},
we proved that this approximation is indeed justified in various cases. Looking back 
to the results of Sects.~\ref{sec:theory} and \ref{sec:isotropic}, we realize that 
the isotropic approximation is equivalent to postulating an ``effective" system of 
equations, 
\begin{equation} \label{effective}
\dot\xi_i   =
\left( \matrix{    0          &    1 \cr
                      -K(t)   &    0          }\right) \xi_i\; ,
\end{equation}
where $\xi_i=(\delta q_i,\delta p_i)$ is the projection of the tangent vector $\xi$
on the subspace of the $i$-th degree of freedom. The equations above represent the
evolution of a typical component of $\xi$. In this sense they could
alternatively be called ``mean-field" or ``single-particle" equations.
The scalar object $K(t)$ is an effective random process which substitutes the Hessian
${\bf V}(t)$, and is in principle unknown.
However, its first two cummulants can be identified in the following way. 
First solve Eq.~(\ref{effective}) 
for the average of $\xi_i^2$ by using second order perturbation theory, as done in 
Sect.~\ref{sec:theory} [just change ${\bf V}$ by $K$, and set $N$=1 in Eq.~(\ref{lambdaM2})]. 
Notice that, as $K(t)$ is a real number, the blocks of the effective 
$\widehat{\bf \Lambda}$ are also real numbers, and the ``isotropic approximation"
is exact now. Then the matrix one must diagonalize to obtain the Lyapunov exponent 
is exactly that of Eq.~(\ref{iso}), provided one makes the identifications:
\begin{eqnarray}
\langle K \rangle & = & \frac{1}{N} {\rm Tr } \langle {\bf V} \rangle  \; ,    \\
\langle \delta K(0) \delta K(\tau) \rangle & = &  
\frac{1}{N} {\rm Tr } \langle \delta {\bf V}(0) \delta {\bf V}(\tau) \rangle  \; ,
\end{eqnarray}
with $\delta K = K -\langle K \rangle$. From this point of view, the isotropic basis
$\{{\bf I}_1,{\bf I}_2,{\bf I}_3\}$ is a single-particle basis. It is the most natural
one in the sense that it treats all degrees of freedom on the same footing.

The perturbative-isotropic approximation, as presented above, is very similar to the 
geometric approach. In fact, in the geometric method \cite{casetti00} an effective equation like 
(\ref{effective}) is proposed, containing the unknown process $K'(t)$. Then, it is argued that 
the first two cummulants of $K'$ are related to the Laplacian of the potential, 
$\triangle {\cal V}\,(t)$:
\begin{eqnarray}
\langle K' \rangle & = & \frac{1}{N} \langle \triangle {\cal V}\, \rangle   \; ,    \\
\langle \delta K'(0) \delta K'(\tau) \rangle & = &  
\frac{1}{N}  \left\langle \left( \delta \triangle {\cal V} \right)^2  
            \right\rangle  \; \bar{\tau} \delta(\tau) \; , \label{deltacorr}
\end{eqnarray}
where 
$\delta \triangle {\cal V}= \triangle {\cal V}- \langle \triangle {\cal V} \, \rangle$, 
and $\bar{\tau}$ is the correlation time of the process $K'(t)$, which is assumed to be 
delta-correlated.

It is obvious that the averages of both processes $K$ and $K'$ coincide because 
${\rm Tr } {\bf V} = \triangle {\cal V}$. So, the differences between the geometric
method and the perturbative-isotropic approach appear only in the fluctuations. 
The similarity between both theories could be enhanced by relaxing the delta-correlation 
assumption of the geometric method, and substituting Eq.~(\ref{deltacorr}) by
\begin{equation}
\langle \delta K'(0) \delta K'(\tau) \rangle =  
\frac{1}{N}  \langle \delta \triangle {\cal V}(0) \delta \triangle {\cal V}(\tau) \rangle \; . 
\end{equation}
But even so, we have not been able to find any analytical relationship between the correlation 
functions of $K$ and $K'$: In principle, the difference between both is non-negligible, and
both effective theories will lead to different estimates for the 
Lyapunov exponent. We expect that numerical simulations will decide which estimate is better.

One comment about the correlation time $\bar{\tau}$ of Eq.~(\ref{deltacorr}) is in order:
Geometric arguments lead to the estimate \cite{casetti00}
\begin{equation} \label{taubar}
\bar{\tau} = \frac{ \pi \sqrt{ \bar{\mu}  }} 
                        { 2 \sqrt{ \bar{\mu} (\bar{\mu}+ \bar{\sigma}) }+ \pi \bar{\sigma}} \; ,
\end{equation}
with $\bar{\mu}\equiv \langle K' \rangle$ and $\bar{\sigma}^2\equiv \langle (\delta K')^2 \rangle$.
Some slightly different expressions have also been proposed \cite{mehra97,barre01,geometric}.
The criterion for testing the accuracy of theses estimates has been the agreement between the geometric
estimate for $\lambda$ and numerical simulations, i.e., the goodness of fit (which is indeed excellent in some
cases). To our knowledge, there is no {\em independent} test of the expression (\ref{taubar}), or others,
in the literature.
Accordingly, a precise definition of $\bar{\tau}$ seems to be lacking. (Is $\bar{\tau}$ equal
to the integral of the normalized autocorrelation function of $\triangle {\cal V}(t)$?)
This is a point which affects the consistency of the geometric method.
Unless a definition is given, to some extent, $\bar{\tau}$ will have the status of a fitting parameter.
Comparisons of the geometric method with other theories will have to take this fact into account.

%%%%%%%%%%%%%%%%%%%%%%%%%%%%

\section{Summary}
\label{sec:conclusions}

%%%%%%%%%%%%%%%%%%%%%%%%%%%%

We showed that the evolution equation in tangent space can be
thought of as a stochastic differential equation with multiplicative noise.
Then, an analytical estimate for the largest Lyapunov exponent of a many-particle system
in equilibrium was derived by using standard perturbative techniques.
Our analysis has been focused on the second-order approximation. 
In this case the Lyapunov exponent can be obtained 
by diagonalizing a matrix whose entries are calculated from the first 
two cummulants of the Hessian of the potential energy, i.e., the averages $\langle V_{ij} \rangle$ and 
the correlation functions $\langle \delta V_{ij}(0) \delta V_{kl}(\tau) \rangle$.
The dimension of this matrix is, in principle, of the order of $N$$\times$$N$,
but we have proposed the conjecture, based on an analogy with the Hermitian problem,
that diagonalization in a truncated basis may be enough to obtain satisfactory results. 

In the crudest approximation, which consists in choosing the three-dimensional isotropic 
basis of Eq.~(\ref{i1i2i3}), the Lyapunov exponent is extracted from a 3$\times$3 matrix.
We argued that this ``isotropic approximation" is equivalent to modeling the tangent dynamics 
of the many-particle system by an ``effective" process $K(t)$ for a single degree of freedom. 
In this way we established a connection with the so-called geometric method, the alternative 
effective theory for estimating the Lyapunov exponent. Both theories are very similar, but differ 
at the point of the definition of the correlation function of $K(t)$. The difference is 
non-trivial and is expected to lead to different predictions.  

In special cases, e.g., one-dimensional lattice-systems with infinite-range 
interactions, we have been able to prove that the isotropic approximation is exact. 
However, in the general case, it may be necessary to consider larger bases. We have 
given examples where these bases are constructed by following the symmetries of the moments 
of ${\bf V}(t)$.

The theory we have presented is perturbative. Loosely speaking, we expect to obtain good 
estimates of Lyapunov exponents in weakly chaotic regimes. 
More quantitatively, the domain of validity of the theory is controlled by the ``Kubo number" $\sigma \tau_c$, 
which quantifies the strength of the fluctuations $\delta {\bf V}(t)$. 
For a given system, it is difficult to say a priori in which regimes the theory will work
satisfactorily. This question and others, like the validity of the isotropic
approximation, and its comparison with the geometric method, will be decided with the
aid of forthcomming numerical simulations.

%%%%%%%%%%%%%%%%%%%%%%%%%%%

\section*{Acknowledgements}

%%%%%%%%%%%%%%%%%%%%%%%%%%%

We are grateful to A. O. Caldeira, C. H. Lewenkopf, F. Nobre, A. M. Ozorio de Almeida,
H. M. Pastawski, A. Robledo, A. M. C. de Souza, and C. Tsallis for fruitful discussions.
We acknowledge Brazilian Agencies CNPq, FAPERJ and PRONEX for financial
support.

%%%%%%%%%%%%%%%%%%%%%%%%%%%

\appendix

\section{The infinite-range case}
\label{apendice}

We have seen in Sect.~\ref{sec:examples} that in the case of a
one-dimensional
system with infinite range interactions the subspace spanned by the matrices
$\hat{\Lambda}^k \openone$ is six-dimensional. An ortoghonal
basis for this subspace is the set
$\{ {\bf I}_1,{\bf I}_2,{\bf I}_3,{\bf Z}_1,{\bf Z}_2,{\bf Z}_3 \}$, where
\begin{equation}
\label{a1}
{\bf Z}_k={\bf Y}_k - {\bf I}_k \; .
\end{equation}
The 6$\times$6 matrix of $\hat{\bf\Lambda}$ can be naturally split into four
blocks of size 3$\times$3.
The block $I$-$I$ has already been calculated [Eq.~\ref{iso}]. Let's now
calculate
the block $I$-$Z$, i.e.,
\begin{equation}
\Lambda_{ij}^{IZ} = \frac
                    { {\rm Tr} \, (\widehat{\bf \Lambda} {\bf Z}_j) {\bf
I}_i}
                    { {\rm Tr} \, {\bf I}_i^2 }   \; .
\end{equation}
By setting ${\bf V}=0$ in Eq.~(\ref{lambdaM2}) we obtain the operator 
$\hat{\bf \Lambda}_0$ ($\hat{\bf \Lambda}$ in the absence of interactions). 
It has the following properties:
\begin{eqnarray}
\hat{\bf \Lambda} {\bf Y}_j & = &  \hat{\bf \Lambda}_0 {\bf Y}_j \; , \\
{\rm Tr} \,  \left( \hat{\bf \Lambda}_0 {\bf Y}_j \right) {\bf I}_i & = &
{\rm Tr} \,  \left( \hat{\bf \Lambda}_0 {\bf I}_j \right) {\bf I}_i \; .
\end{eqnarray}
Using these two properties together with Eq.~(\ref{a1}) one arrives at
\begin{equation}
\Lambda_{ij}^{IZ} =  \Lambda_{0,ij}^{II} - \Lambda_{ij}^{II} \; .
\end{equation}
Analogously one obtains
\begin{eqnarray}
\Lambda_{ij}^{ZI} & \approx &   \frac{1}{N} \Lambda_{ij}^{IZ} \; , \\
\Lambda_{ij}^{ZZ} & \approx &   \Lambda_{0,ij}^{II} +  \frac{1}{N}
\Lambda_{ij}^{II} \; ,
\end{eqnarray}
where the symbol $\approx$ means that terms of relative size $1/N$ have 
been discarded. The 6$\times$6 matrix reads
\begin{equation}
{\bf \Lambda} \approx
\left( \matrix{ {\bf \Lambda}^{II}
&
               - {\bf \Lambda}^{II}_1
\cr
    - \frac{1}{N}{\bf \Lambda}^{II}_1
&
                {\bf \Lambda}^{II}_0  + \frac{1}{N}{\bf \Lambda}^{II}_1
\cr
  } \right)                                \; ,
\end{equation}
with the definition 
${\bf \Lambda}^{II}_0 + {\bf \Lambda}^{II}_1  = {\bf \Lambda}^{II}$.
Then it can be checked that the matrix above has three zero eigenvalues while  
the remaining three are the eigenvalues of the matrix
\begin{equation}
{\bf \Lambda}^{II}_0  + \frac{N+1}{N}{\bf \Lambda}^{II}_1 \approx  {\bf
\Lambda}^{II}  \; .
\end{equation}
Thus the isotropic approximation is essentially exact for the infinite-range
$XY$-Hamiltonian.

%%%%%%%%%%%%%%%%%%%%%%%%%%%

% BIBLIOGRAPHY

\end{multicols}

\end{document}